\documentclass[twocolumn]{el-author}

\usepackage{algorithm}
\usepackage{algorithmic}
\usepackage{subfigure}

\newcommand{\ua}{\uparrow}

\newcommand{\nc}{\newcommand}

\nc{\da}{\downarrow} \nc{\hc}{\hat{c}} \nc{\hS}{\hat{S}}

\nc{\bra}{\langle} \nc{\ket}{\rangle} \nc{\eq}{equation (\ref}

\nc{\h}{\hat} \nc{\hT}{\h{T}}\nc{\be}{\begin{eqnarray}}

\nc{\ee}{\end{eqnarray}}\nc{\rd}{\textrm{d}}\nc{\e}{eqnarray}\nc{\hR}{\hat{R}}\nc{\Tr}{\mathrm{Tr}}

\nc{\tS}{\tilde{S}}\nc{\tr}{\mathrm{tr}}\nc{\8}{\infty}\nc{\lgs}{\bra\ua,\phi|}\nc{\rgs}{|\ua,\phi\ket}

\nc{\hU}{\hat{U}}\nc{\lfs}{\bra\phi|}\nc{\rfs}{|\phi\ket}\nc{\hZ}{\hat{Z}}\nc{\hd}{\hat{d}}\nc{\mD}{\mathcal{D}}

\nc{\bd}{\bar{d}}\nc{\bc}{\bar{c}}\nc{\mc}{\mathcal}\nc{\ea}{eqnarray}\nc{\mG}{\mathcal{G}}\nc{\bce}{\begin{center}}

\nc{\ece}{\end{center}}

\date{11th August 2018}

\begin{document}

\title{High-precision timing and frequency synchronization method for MIMO-OFDM systems in double-selective channels}

\author{Jun Liu, Kai Mei, Xiaochen Zhang, Xiaoying Zhang, Dongtang Ma, Jibo Wei}

\abstract{In this letter, a novel synchronization method for MIMO-OFDM systems is proposed. The new approach has an accurate estimate of both symbol timing and large frequency offest. Simulation results show the excellent robustness of our method in double-selective channel even if the strongest multipath component arrives behind the first path.}

\maketitle

\section{Introduction}

The accurate timing and frequency synchronization is a crucial prerequisite for high-speed and reliable wireless communication \cite{1}. In many scenarios, either transceiver moves in complex terrain at a high speed. Under these circumstances, wireless channels are usually double-selective and have large delay paths, which may degrade performance of whole system significantly. In order to tackle this problem, we proposes a CAZAC (Const Amplitude Zero Auto-Corelation) sequence aimed high-precision synchronization algorithm for multiple-input multiple-output orthogonal frequency division multiplexing (MIMO-OFDM) systems.

\section{Preamble structure and system model}
As shown in Fig.1, one preamble is composed by 4 OFDM symbols (${N_{{\rm{FFT}}}=512}$, ${N_{{\rm{CP}}}=128}$) which are generated by two CAZAC Zadoff-Chu sequences. 

\begin{figure}[h]
\label{fig:1}
\centering{\includegraphics[width=76mm]{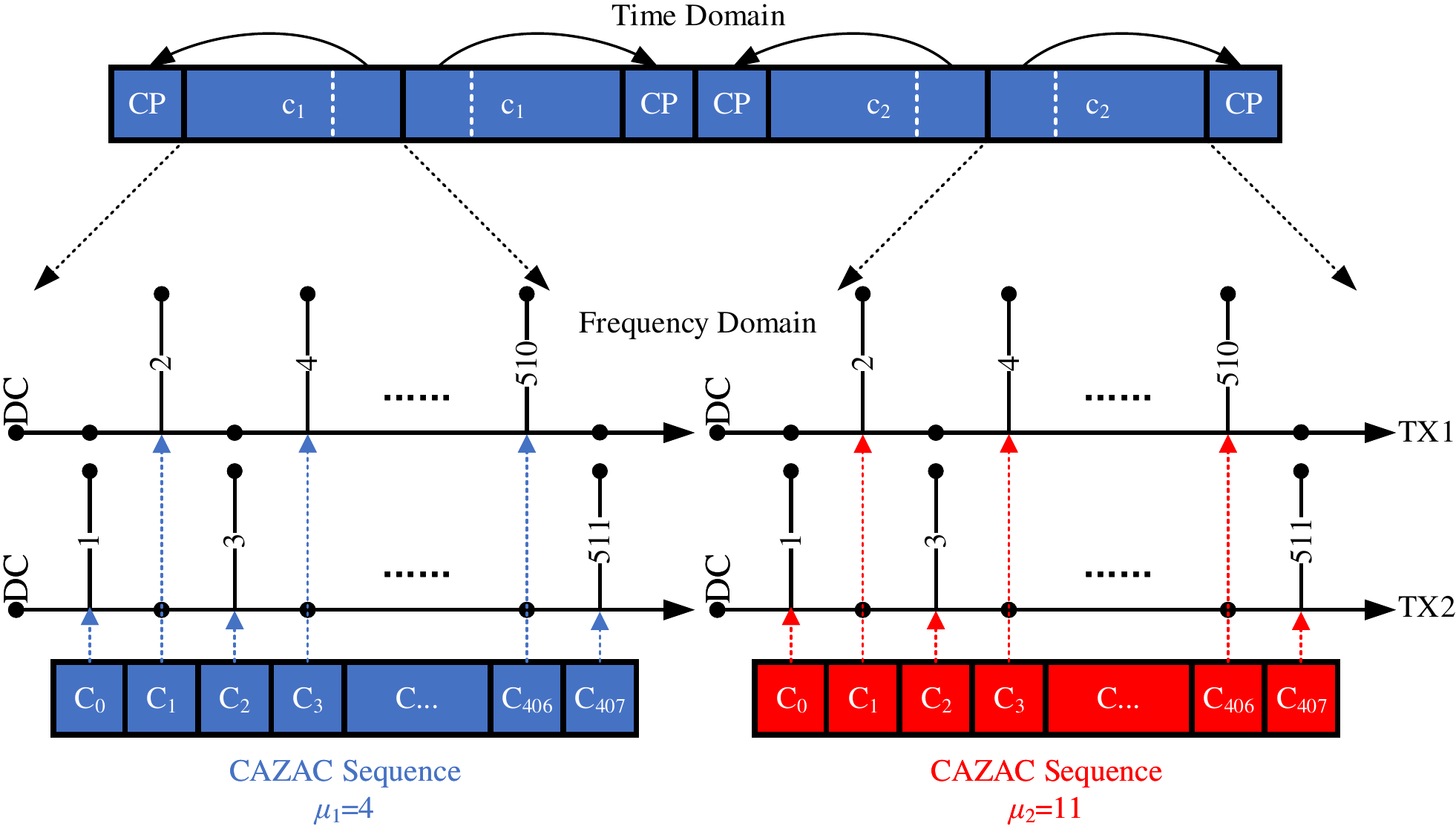}}
\caption{Frequency-domain-orthogonal preamble structure for MIMO-OFDM system with two transmit antennas
\source{}}	
\end{figure}

Orthogonal structure is adopted in frequency domain. Specifically, each CAZAC sequence is separated into two equal-length parts and distributed on even and odd subcarriers of TX1 and TX2, respectively. CAZAC sequence is defined as
\begin{align}
{C_{{\mu _i}}}\left[ k \right] = \left\{ {\begin{array}{*{20}{c}}
{\exp \left( { - j2\pi {k^2}{\mu _i}/\left( {2{N_\mu }} \right)} \right), {\kern 1pt} {\kern 1pt} {\kern 1pt} {\kern 1pt} {\kern 1pt} {\kern 1pt} {\kern 1pt} {\kern 1pt} {\kern 1pt} {\kern 1pt} {\kern 1pt} {\kern 1pt} {\kern 1pt} {\kern 1pt} {\kern 1pt} {\kern 1pt} {\kern 1pt} {\kern 1pt} {\kern 1pt} {\kern 1pt} {\kern 1pt} {\kern 1pt} {\kern 1pt} {\kern 1pt} {\kern 1pt} {\kern 1pt} {\kern 1pt} {N_\mu }{\kern 1pt} {\kern 1pt} {\kern 1pt} {\kern 1pt} {\kern 1pt} {\rm{even}}}\\
{\exp \left( { - j2\pi k\left( {k + 1} \right){\mu _i}/\left( {2{N_\mu }} \right)} \right),{\kern 1pt} {\kern 1pt} {\kern 1pt} {\kern 1pt} {\kern 1pt} {\kern 1pt} {\kern 1pt} {N_\mu }{\kern 1pt} {\kern 1pt} {\kern 1pt} {\kern 1pt} {\kern 1pt} {\rm{odd}}}
\end{array}} \right.
\end{align}
where $N_\mu$ is the sequence length, $k$ is the signal index and $\mu_i$ is the parameter of CAZAC sequence.
The second and fourth symbols are duplicate of the first and third symbols. DC and virtual subcarriers are unoccupied. The repetitive pattern in the preamble could enhance the performance of frame synchronization at the beginning. Besides, this reiterative structure is necessary for fraction CFO (Carrier Frequency Offset) estimation since the phase rotation between two identical symbol is a measure for the frequency offset.

Without loss of generality, we assume that the MIMO-OFDM system has 2 antennas at both TX and RX. Our approach could be easily extend to more complex systems. Here, the transmitted signal and received signal at a discrete time instance $n$ are defined as an $2\times1$ complex vector ${\bf{c}}\left(n\right)$ and ${\bf{r}}\left(n\right)$, respectively. The channel impulse response (CIR) of double-selectccive channel is represented as ${{\bf{H}}\left( {\tau ,n} \right)}$ and its $(q,p)$ element ${h_{qp}}\left( {\tau ,n} \right)$ describes the channel gain between $p$th transmitter antenna and $q$th receiver antenna with delay $\tau$ at time instance $n$. Then, the discrete-time MIMO-OFDM baseband signal model is given by 
\begin{align}
{\bf{r}}\left( n \right) = \sum\limits_{\tau  = 0}^{L - 1} {{\bf{H}}\left( {\tau ,n} \right)} {\bf{c}}\left( {n - \tau } \right) + {\bf{w}}\left( n \right)
\end{align}	
where ${\bf{w}}\left( n \right)$ represents additive white Gaussian noise (AWGN). The time domain signal ${\bf{c}}(t)$ is generated by IDFT:
\begin{align}
c_l^p\left( n \right) = \frac{1}{{\sqrt {{N_{{\rm{FFT}}}}} }}\sum\limits_{k = 0}^{{N_{{\rm{FFT}}}} - 1} {X_l^p\left[ k \right]} \exp \left( {j2\pi kn/{N_{{\rm{FFT}}}}} \right)	
\end{align}
where ${X_l^p\left[ k \right]}$ represents the data on $k$th sub-carrier in $l$th symbol deploying in $p$th antenna.

\section{Coarse symbol timing offest estimation}
The purpose of coarse symbol timing offset (STO) estimation is to detect the arrival of preamble. Delay correlation criterion is proposed and given by
\begin{align}
{\rm{ST}}{{\rm{O}}_{{\rm{coarse}}}}{\rm{ = }}\mathop {{\rm{argmax}}}\limits_n  \left[ {\sum\limits_{m =  - {N_{{\rm{CP}}}}}^{{N_{{\rm{CP}}}} - 1} {\frac{{\sum\limits_{p = 1}^{{N_t}} {{{\left| {\Lambda \left( {{n_p}} \right)} \right|}^2}} }}{{\sum\limits_{p = 1}^{{N_t}} {P{{\left( {{n_p} - {N_{{\rm{FFT}}}}} \right)}^2} + P{{\left( {{n_p}} \right)}^2}} }}} } \right]
\end{align}
where ${n_p} = n - \left( {{N_t} - p} \right){N_{{\rm{train}}}} + m $, ${N_{{\rm{train}}}}=2\left( {{N_{{\rm{FFT}}}}+{N_{{\rm{CP}}}}} \right)$, $\Lambda \left( n \right)$ and $P\left( n \right)$ are defined as
\begin{align}
\Lambda \left( n \right) = \sum\limits_{i = n - \left( {{N_{{\rm{FFT}}}} - 1} \right)}^n {\sum\limits_{q = 1}^{{N_r}} {r_q^ * \left( {i - {N_{{\rm{FFT}}}}} \right){r_q}\left( i \right)} }	
\end{align}
and
\begin{align}
P\left( n \right) = \sum\limits_{i = n - \left( {{N_{{\rm{FFT}}}} - 1} \right)}^n {\sum\limits_{q = 1}^{{N_r}} {r_q^ * \left( i \right){r_q}\left( i \right)} }
\end{align}
respectively.
Practically, setting threshold is unreasonable for hardware implementation due to the influences of channel, noise as well as automatic gain control (AGC). Therefore, $2{N_{{\rm{CP}}}}$-length moving average is necessary.

\section{Carrier frequency offset estimation}
The relationship between estimation range of normalized CFO $\hat \varepsilon$ and the length of periodical structure $o$ is
\begin{align}
\hat \varepsilon  \in \left( { - \frac{{{N_{{\rm{FFT}}}}}}{{2o}},\frac{{{N_{{\rm{FFT}}}}}}{{2o}}} \right)
\end{align}
the range of fractional CFO estimator is $\left({-0.5,0.5}\right)$. To overcome this limitation, an algorithm is proposed for joint estimation of fractional and integral CFO algorithm. This algorithm is summarized in Algorithm \ref{alg:1}.

\begin{algorithm}
\caption{The joint estimation algorithm for fractional and integral CFO}
\label{alg:1}
\begin{algorithmic}[1]
\STATE {{\bf{Estimate}} fractional CFO by \\

${\hat \varepsilon _{{\rm{fraction}}}} = {\left. {\arg \left[ {\sum\limits_{p = 1}^{{N_t}} {\Lambda \left( {{n_p}} \right)} } \right]/\left( {2\pi } \right)} \right|_{n = {\rm{ST}}{{\rm{O}}_{{\rm{coarse}}}}}}$
}
\STATE {{\bf{Compensate}} receive signal by ${{\hat \varepsilon }_{{\rm{fractional}}}}$} temporarily:\\
${\bf{r'}}\left( n \right) \leftarrow {\bf{r}}\left( n \right)\exp \left( {-j2\pi n{{\hat \varepsilon }_{{\rm{fractional}}}}/{N_{{\rm{FFT}}}}} \right) $

\STATE {{\bf{Estimate}} integral CFO by\\
${{\hat \varepsilon }_{{\rm{intgeral}}}} = \mathop {\arg \max }\limits_n \left\{ {{\rm{dot\_product}}\left[ {I\left( n \right),c_{{\mu _1} - {\mu _2}}^ * } \right]} \right\} - {\rm{ST}}{{\rm{O}}_{{\rm{coarse}}}} $

$I\left( n \right) = {\rm{conv}}\left\{ {\sum\limits_{q = 1}^{{N_r}} {{r '_q}\left( n \right)} ,\left[ \sum\limits_{q = 1}^{{N_r}} {{\rm{fliplr}}\left[ {{r '_q}\left( {n + {N_{{\rm{train}}}}} \right)} \right]} \right]^* } \right\} $
}

\STATE{
${\hat \varepsilon }= {{\hat \varepsilon }_{{\rm{fractional}}}} + {{\hat \varepsilon }_{{\rm{integral}}}} $
}

\STATE {{\bf{Compensate}} receive signal by $-{{\hat \varepsilon }}$}:\\
${\bf{r''}}\left( n \right) \leftarrow {\bf{r}}\left( n \right)\exp \left( {-j2\pi n{{\hat \varepsilon }}/{N_{{\rm{FFT}}}}} \right) $ 
\end{algorithmic}	
\end{algorithm}

In step 3, we utilize the property of CAZAC sequence(${c_{{\mu _1}}} \cdot c_{{\mu _2}}^ *  \to {c_{{\mu _1} - {\mu _2}}} $) and Fourier transform($f\left( { - t} \right) \leftrightarrow F\left( { - j\omega } \right) $, ${f^ * }\left( t \right) \leftrightarrow {F^ * }\left( { - j\omega } \right) $). Operators "$*$", ${\rm{fliplr}}\left[  \cdot  \right]$, and ${\rm{conv}}\left[ {{\bf{A}}, {\bf{B}}} \right]$ represent conjugate operator, reversing a sequence and convolving $N_{\rm{FFT}}$-length (from instant $n$ to $n+ N_{\rm{FFT}}-1 $) vectors {\bf{A}} and {\bf{B}}. The search window should be centered around ${\rm{ST}}{{\rm{O}}_{{\rm{coarse}}}}$. It is necessary to explain that why receive signal need to be temporarily compensated by offset $-{{\hat \varepsilon }_{{\rm{fractional}}}} $ in step 2. If $\left|{\varepsilon}\right|<0.5 $, the fractional CFO estimator will function well and the integral CFO will be estimated as 0 by using ${\bf{r'}}(n)$. On the contrary, if $\left|{\varepsilon}\right|\ge0.5 $, the joint estimation mechanism is more complex and it can be represented as Fig.2.

\begin{figure}[h]
\label{fig:2}
\centering{\includegraphics[width=85mm]{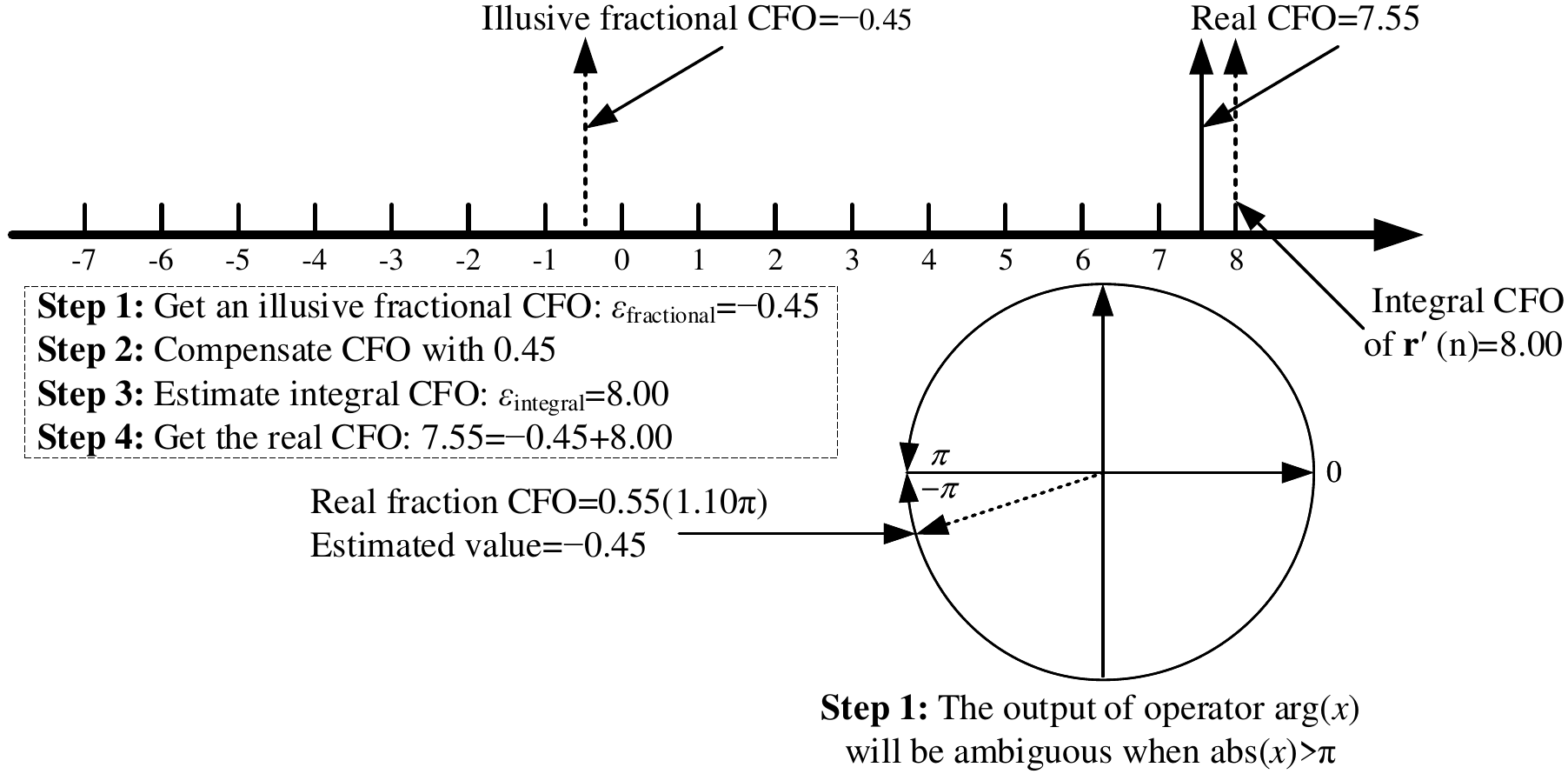}}
\caption{The mechanism of joint CFO estimation algorithm when $\left|{\varepsilon}\right|\ge0.5 $
\source{}}	
\end{figure}
For example, in step 1, due to the ambiguity of operator $\arg \left(  \cdot  \right) $, if CFO=7.55, estimated fractional CFO will be $-0.45$ instead of 0.55. Then, we will get estimated integral CFO value $8.00$ by using temporary compensated signal ${\bf{r'}}(n)$. Finally, the real CFO will be acquired by adding the fractional CFO to integral CFO.

\section{Fine symbol timing synchronization}
Once the CFO is acquired, fine STO estimation can be carried out and represented in the form of cross correlation between the compensated signal and local CAZAC sequences. However, performance of synchronization is undermined significantly when the strongest path is not the ahead of all other paths. In order to find the exact timing symbol, we proposed a means which is summarized in Algorithm 2.

\begin{algorithm}
\caption{Robust fine timing synchronization algorithm}
\label{alg:2}
\begin{algorithmic}[1]
\STATE {According a presetted false alarm probability $P_{\rm{FA}}$ to set threshold TH}
\FOR{k=1:K}

\STATE {{\bf{Calculate}} cross correlation measure:\\
${\rm{temp = }} \mathop {{\rm{argmax}}}\limits_n \left[ {X\left( n \right)/N\left( n \right)} \right] $}\\

$X\left( n \right) = {\left| {{\rm{sum}}\left[ {{R_1}\left( {{n_1}} \right)} \right]{\rm{ + sum}}\left[ {{R_2}\left( {{n_2}} \right)} \right]} \right|^2} $\\

$N\left( n \right) = E \sum\limits_{m = 0}^{{N_{{\rm{FFT}}}}} {\left[ {{{\left| {\sum\limits_{q = 1}^{{N_r}} {{{{\rm{r''}}}_q}\left( {{n_1} + m} \right)} } \right|}^2} + {{\left| {\sum\limits_{q = 1}^{{N_r}} {{{{\rm{r''}}}_q}\left( {{n_2} + m} \right)} } \right|}^2}} \right]} $\\
where $E$ is the total energy of $c_1$ and $c_2$.

${R_1}\left( {{n_1}} \right) = {\rm{conv}}\left\{ {\sum\limits_{q = 1}^{{N_r}} {{{r''}_q}\left( {{n_1}} \right), \left[ {\rm{fliplr}}\left[ {{c_1}} \right] \right]^*} } \right\} $\\

${R_2}\left( {{n_2}} \right) = {\rm{conv}}\left\{ {\sum\limits_{q = 1}^{{N_r}} {{{r''}_q}\left( {{n_2}} \right), \left[{\rm{fliplr}}\left[ {{c_2}} \right] \right]^*} } \right\} $\\

${n_1} = n + {\rm{ST}}{{\rm{O}}_{{\rm{coarse}}}} + 0.5{N_{{\rm{train}}}} $\\

${n_2} = n + {\rm{ST}}{{\rm{O}}_{{\rm{coarse}}}} + {N_{{\rm{train}}}} + {N_{{\rm{CP}}}} $

\IF{$X\left( \rm{temp} \right)/N\left( \rm{temp} \right) > {\rm{TH}} $}
\STATE {record \rm{temp} into  array ${\bf{D}} $}
\ENDIF

\ENDFOR

\STATE{${\rm{ST}}{{\rm{O}}_{{\rm{fine}}}} = \min \left( {\bf{D}} \right) $}

\end{algorithmic}	
\end{algorithm}

Both the threshold TH and K are crucial for localizing precise timing symbol against the effects from channel and noise. According to central limit theorem, $\sqrt{X(n)/N(n)}$ can be modeling as Rayleigh stochastic process and then TH is given by
\begin{align}
{\rm{TH}} = \sqrt {- \ln \left( {{P_{{\rm{FA}}}}} \right) \cdot 2\sigma _0^2}	
\end{align}
where ${\sigma_0}=\frac{{\sqrt {2/\pi } }}{{\left\langle {I'} \right\rangle }} \sum\limits_{I'} {\sqrt {X(n)/N(n)}}$. $I'$ is the set of time indices that is complementary to the set of indices $I = \left[ {{\rm{ST}}{{\rm{O}}_{{\rm{coarse}}}} - 5{\rm{K}},{\rm{ST}}{{\rm{O}}_{{\rm{coarse}}}} + 5{\rm{K}}} \right]$ and ${{\left\langle {I'} \right\rangle }} $ denotes the cardinality of the set $I'$. Practically, K=3 or 4 and $P_{\rm{FA}}=10^{-8} $ are appropriate.


\section{Simulation results}
The performance of the proposed technique deployed in $2\times2$ SFBC MIMO-OFDM system has been evaluated through simulation. Comparison with two state-of-the-art methods (Schenk's\cite{2} and Park's methods\cite{3,4}) has made. Mobile speed is set to 120km/h and make CFO=7.55 for proposed method (CFO=0.45 for other two methods due to the limitation of estimation range). Other simulation parameters are defined in Table 1. 
\begin{table}[htp]
\label{tab:1}
\centering
\caption{Simulation Parameters}
\begin{tabular}{|c|c|}
\hline
System Parameter           & Parameter Value \\ \hline
Modulation                 & QPSK            \\ \hline
Bandwidth                  & 8MHz            \\ \hline
Number of data subcarriers & 408             \\ \hline
Code Rate                  & 0.5 (Turbo)     \\ \hline
Channel                    & Vehicular A\cite{5}     \\ \hline
\end{tabular}
\end{table}

\begin{figure}[htp]
\centering
\subfigure[Timing error]{\label{fig:3a}
\includegraphics[width=43mm]{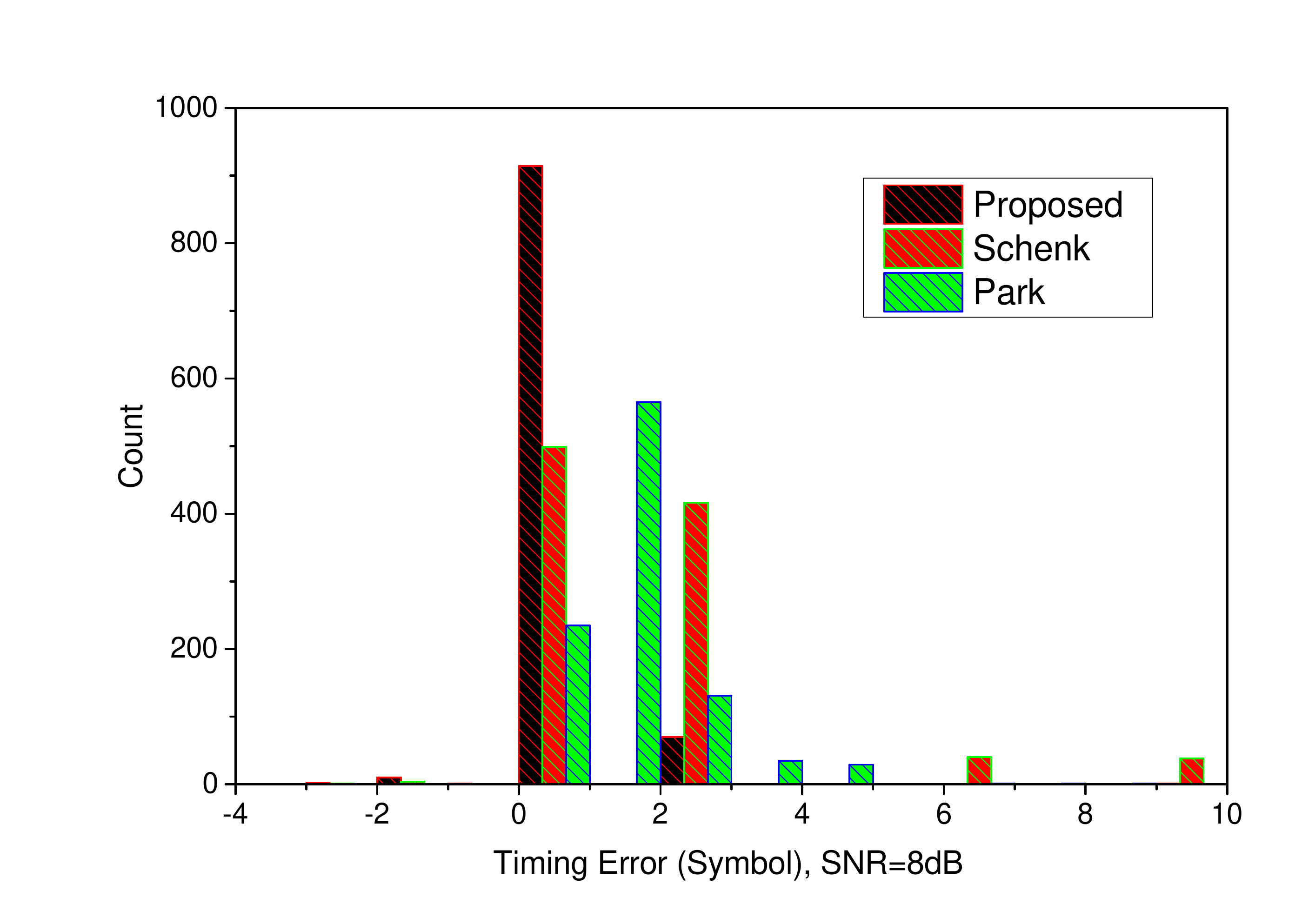}}
\subfigure[MSE of CFO estimation]{\label{fig:3b}
\includegraphics[width=41.5mm]{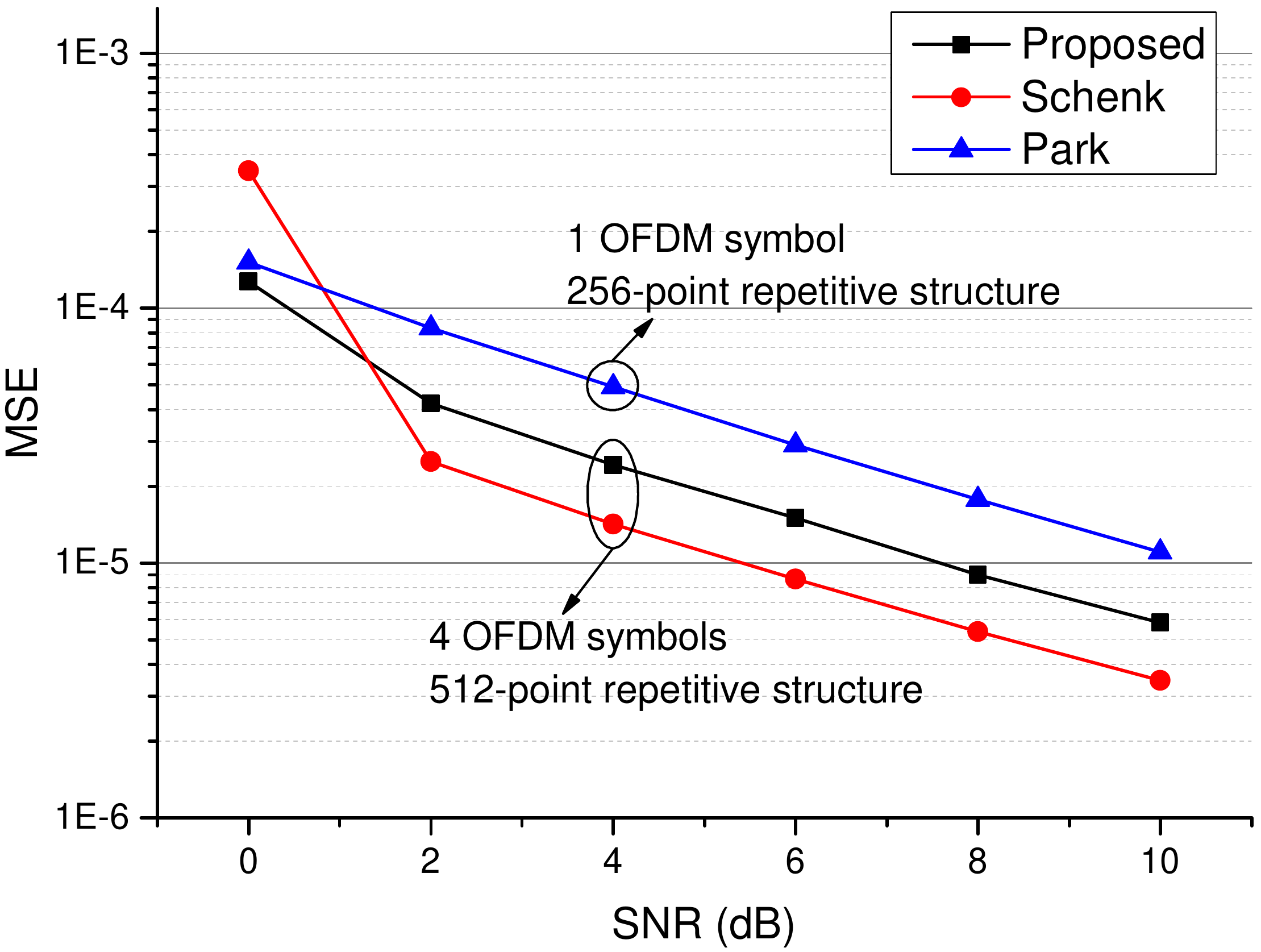}}
\caption{Synchronization performance}
\label{fig:3}
\end{figure}

Fig.3 shows the performance of three kinds of synchronization methods. It can be seen proposed method is the most stable timing approach and Park's method has the worst performance. Bit error rate (BER) curve is illustrated in Fig.4. From this perspective, three methods are performing similarly though they have slight performance differences in CFO estimation.

\begin{figure}[h]
\label{fig:4}
\centering{\includegraphics[width=50mm]{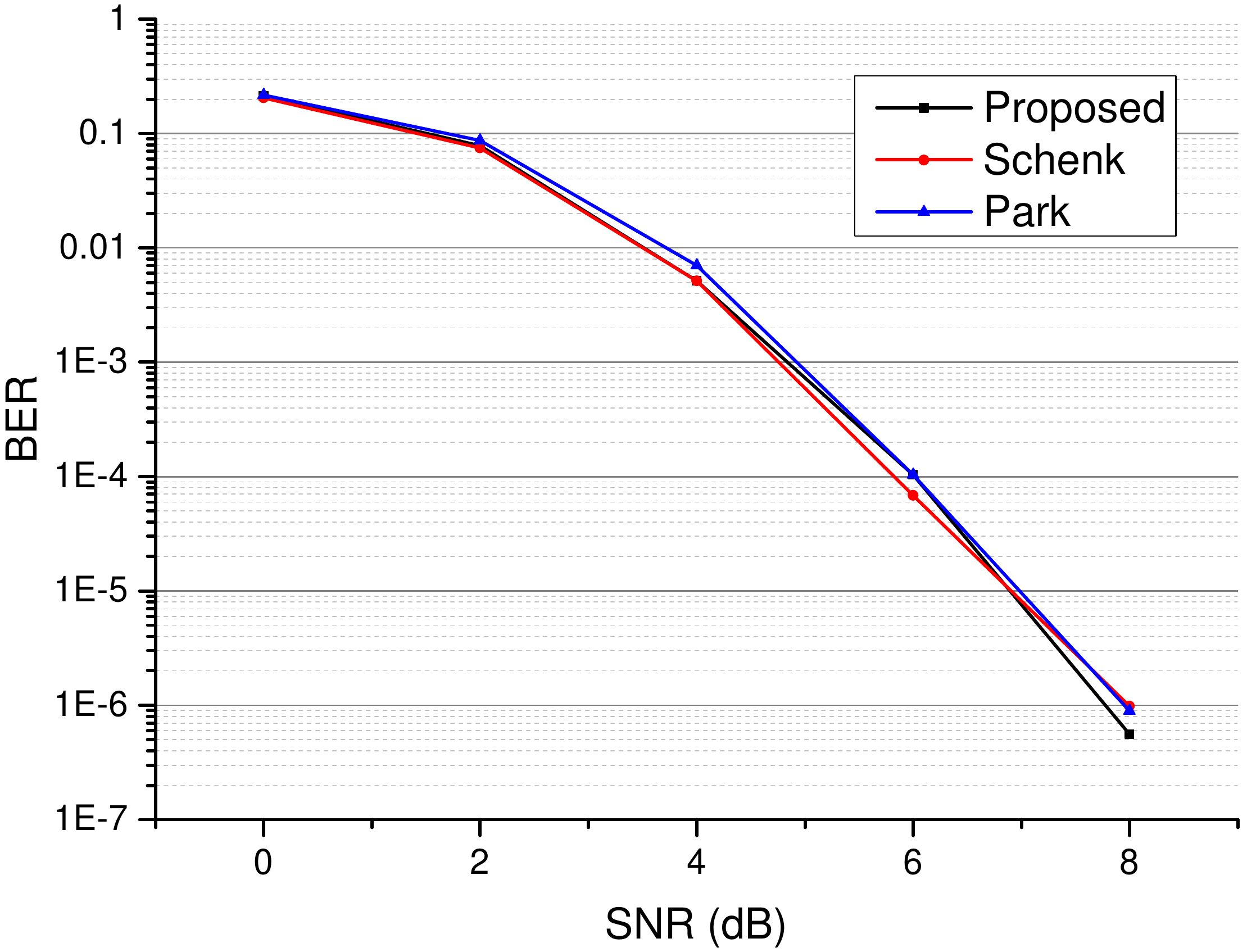}}
\caption{BER performance result with LS channel estimation method
\source{}}	
\end{figure}

\section{Conclusion}

In this letter, we propose a high-precision synchronization  scheme for MIMO-OFDM system. This scheme shows excellent performance comparing with other two state-of-the-art methods. Simulation verifies that the proposed scheme could be applied in high-speed and large-delays scenarios. Moreover, the proposed scheme can be easily extended to systems with multiple antennas.
\vskip3pt

\ack{This work was supported by the Research Found of National University of Defense Technology under Grant ZK17-03-13.}


\vskip5pt

\noindent Jun Liu et al. (\textit{BCNG, National University of Defense Technology, Changsha, People's Republic of China})

\vskip3pt

\noindent E-mail: liujun15@nudt.edu.cn


\begin{thebibliography}{}



\bibitem{1}

Nasir A A, et al.: `Timing and carrier synchronization in wireless communication systems: a survey and classification of research in the last 5 years', \textit{Eurasip Journal on Wireless Communications \& Networking}, 2016, \textbf{2016}, (1), pp. 180



\bibitem{2}

Van Zelst A, Schenk T C W.: `Implementation of a MIMO OFDM-based wireless LAN system', \textit{IEEE Trans Signal Processing}, 2004, \textbf{52}, (2), pp. 483-494

\bibitem{3}

Park B, Cheon H, Kang C, et al.: `A novel timing estimation method for OFDM systems', \textit{IEEE Communications Letters}, 2003, \textbf{7}, (5), pp. 239-241


\bibitem{4}

Liu G, Ge J H, Guo Y: `Time and frequency offset estimation for distributed multiple-input multiple-output orthogonal frequency division multiplexing systems', \textit{Iet Communications}, 2010, \textbf{4}, (6) pp. 708-715


\bibitem{5}

`Guidelines for the evaluation of radio transmission technologies for IMT-2000', \textit{Recommendation ITU-R M.1225}, 1997, pp. 28



\end{thebibliography}
\end{document}